\documentclass{elsart} 
 \begin{document}
 \begin{frontmatter}
 \input epsf
 \title{Phase Transitions in Load Transfer Models of Fracture.}
 \author{Y. Moreno$^{1,2}$, J. B. G\'{o}mez$^3$, A.
F. Pacheco$^2$} 
 \address{ $^1$ The Abdus Salam International Centre for Theoretical 
Physics, P.O. Box 586,\\ Strada Costiera 11, I34100, Trieste, Italy.\\$^2$ Departamento 
de F\'{\i}sica Te\'{o}rica, Universidad de Zaragoza, 50009 Zaragoza, Spain.\\ $^3$ 
Departamento de Ciencias de la Tierra, 
 Universidad de Zaragoza, 50009 Zaragoza, Spain.}
\begin{abstract} The possible paralelism existing between phase transitions and fracture 
in disordered materials, is discussed using the well-known Fiber Bundle Models and a 
probabilistic approach suited to smooth  fluctuations near the critical point. Two 
limiting cases of load redistribution are analyzed: the global transfer scheme, and the 
local transfer rule. The models are then studied and contrasted by defining the branching 
ratio as an order parameter indicative of the distance of the system to the critical 
point. In the case of long range interactions, i.e., the global rule, the results 
indicate that fracture can be seen as a second-order phase transition, whereas for the 
case of short range interactions (the local transfer rule) the bundle fails suddenly with 
no prior significant precursory activity signaling the imminent collapse of the system, 
this case being a first-order like phase transition. 
 \end{abstract}
\end{frontmatter}

\section{Introduction}
\label{intro}

Fracture phenomena have attracted a lot of interest in the last several years, both 
experimentally and theoretically. In the lab, a disordered material subjected to an 
increasing external load can be experimentally studied by measuring the acoustic 
emissions before the global rupture. It has been shown \cite{ggbc97,ppvac94} that this 
intense precursory activity in the form of bursts of different microscopic sizes follows 
a well-defined power law. From the theory side, the understanding of fracture in 
heterogeneous materials has progressed due to the use of lattice models and large scale 
numerical simulations \cite{cb97}. Recently, the introduction of models of material 
failure has lead to the evidence that rupture can be viewed as a kind of critical 
phenomenon \cite{cb97,hr90,zrsv97,asl97}.  Nevertheless, the question of whether rupture 
exhibits the properties of a first-order or a second-order phase transition remains under 
discussion as well as what is the order parameter that determines the type of transition. 

In this field, it is important to use models able to describe the complexity of the 
rupture process, although they should be simple enough to permit analytical insights. To 
this class of models belong the well-known Fiber Bundle Models (FBM) widely used since 
their introduction more than forty years ago \cite{c58,d45}. In static FBM, a set of 
fibers (elements) is located on a supporting lattice and one assigns to its elements a 
random strength threshold sampled from a probability distribution. The lattice is loaded 
and fibers break if their loads exceed their threshold values. Now, one can assume 
different load transfer rules to mimic the range of interactions among the fibers in the 
set. The Global Load Sharing rule is the simplest theoretical approach one can adopt to 
make the problem analytically tractable, which implies that the load carried by failed 
elements is equally distributed among the surviving elements of the system, representing 
in this way a long-range interaction among the constituent parts of
the system. This is a kind of mean field approach to the more 
extreme Local Load Sharing scheme, where the interaction among the elements is 
short-range and the load supported by the failing elements is transferred only to a neighborhood. 

Very recently \cite{mgpprl00}, we have developed a probabilistic approach suited to 
smooth the fluctuations around the point of final collapse. In this paper, we extend the 
method to analyze the local load sharing scheme and contrast the two limiting cases of 
range interactions: the long range and the short range.  While for the case of global 
load sharing, we obtain different scaling relations that point out that the fracture of a 
fiber set with long range interactions may be a continuous transition, for the case of 
short range interactions, the failure exhibits no significant precursory activity and the 
system breaks abruptly showing, in this case, the properties of a first order transition 
where no scaling laws can be identified. The rest of the paper is organized as follow. In 
Section\ \ref{fbmmodel}, we present the standard approach to the FBM. In Section\ 
\ref{PAFBM}, we introduce the new probabilistic method and the results derived from it 
and from Monte Carlo simulations. Finally, Section\ \ref{discon} is devoted to discussion 
and conclusions. 

\section{Static Fiber Bundle Models}
\label{fbmmodel}

Let us first recall the basic ingredients of the static FBM and how one proceeds in 
numerical simulations. The term static means that time plays no role in the model 
\cite{vgmp99}. The system under consideration is a set of $N_0$ elements (fibers) located 
in a supporting lattice each one having at the initial state a zero load and a {\it 
fixed} strength threshold value (quenched disorder) sampled randomly from a probability 
distribution $P(\sigma)$. The system is then subjected to an external force $F$ which is 
shared democratically among the fibers, that is, the distribution of the total force 
produces that each element increases its load, $\sigma$, in the same amount. This 
individual stress acts as the control parameter. The loading process is done 
quasistatically; i.e., the external force is increased at a sufficiently slow rate as to 
produce a single breaking event when the stress in the weakest element equals its 
threshold value. Then, the increase of external force stops and the load of the broken 
fiber is transferred according to the transfer scheme assumed. In the global load sharing 
FBM, this implies that the load on fiber $i$ is given by $\sigma_i=F/n_s(F)$, where $n_s$ 
is the number of surviving elements for a given external load $F$. In the local load 
sharing FBM, the above relation does not hold any more because there the redistribution 
of load is performed among the nearest neighbors of the failed fiber giving rise to the 
appearance of regions with different stress concentrations. 

In both schemes, the rupture of a fiber may induce secondary failures which in turn may 
trigger more failures. This process of induced failure at constant external load, termed 
avalanche, stops when all surviving elements carry a load lower than their thresholds. 
The system is then loaded again and the process is repeated until the final catastrophic 
avalanche provokes the total rupture of the material, which occurs at a critical load 
$\sigma_c$ that depends on the probability distribution from where the individual 
strengths were drawn as well as on the system size. At this point, it is worth recalling 
that the exact value of $\sigma_c$ can be analytically obtained in the thermodynamic 
limit for the global load sharing case while for local load sharing schemes there is no 
theoretical approach leading to $\sigma_c$. It is also known that the critical load 
$\sigma_c$ for the global load sharing FMB is non-zero and independent of $N_0$ for 
$N_0\rightarrow\infty$, whereas $\sigma_c$ tends to zero as $N_0$ goes to infinity in the 
case of local load sharing models \cite{d45,hp78}. Fiber bundle models have also been 
recently used in self-organized criticality (SOC), a theoretical framework widely used 
for the study of avalanche phenomena in disordered systems. It has been shown using these 
models that systems with plastic behavior can reach a SOC state just before the global 
rupture \cite{kzh99}. A second case of self-organization with power law distributions in 
several quantities corresponds to the situation in which the fracture process coexists 
with a healing process \cite{mgp99}. 

In numerical simulations, the cycle of complete breakdown of the material is performed 
many times in order to average out the effect of fluctuations. Nevertheless, the stress 
history of a particular element is made of steps, so that the fluctuations around the 
critical load can never be completely avoided. Let fiber $k$ be supporting a stress 
$\sigma_k^0$ at a given step of the failure process. It will continue to support 
$\sigma_k^0$ until a global driving or a stress transfer from one (or more) of the failed 
elements occurs. At this moment, element $k$ instantaneously changes its load to 
$\sigma_k^1=\sigma_k^0+\sigma_{dist}$, where $\sigma_{dist}$ is the
load coming from the 
failed fibers or from the global driving. So, in a later step, element $k$ could receive 
again load, suffering a second step-like increase in stress. This step-like stress 
history continues until fiber $k$ fails. As we are interested in studying the behavior of 
the system as the critical point is approached, it is of utmost importance to find a 
simple method able to capture the evolution of the system near the critical point 
avoiding as much as possible the fluctuations. 

\section{Probabilistic Approach to Static Fiber Bundle Models}
\label{PAFBM}

Let there be a set of $N_0$ elements located on a supporting lattice. Suppose that each 
element carries a given load $\sigma$, which is set to zero at the initial state. Fibers 
break depending on their strengths which are distributed according to a probability 
distribution $P(\sigma)$. Different probability distributions can be considered. In 
materials science the Weibull distribution is widely used, 
 \begin{equation} 
P(\sigma)=1-e^{-(\frac{\sigma}{\sigma_{0}})^{\rho}},
 \label{eq1} 
 \end{equation}
where $\rho$ is the so-called Weibull index, which controls the degree of 
disorder in the system (the bigger the Weibull index, the smaller the disorder), and 
$\sigma_{0}$ is a load of reference. In the following, we will assume $\sigma_{0}=1$, and 
therefore the loads are dimensionless. Although the results we show hereafter are for the 
Weibull distribution to gain in definiteness, they have been also obtained for a wide 
class of distributions. Besides, for continuous distributions decaying fast enough, the 
approach to the critical state does not depend on the detailed form of the disorder 
\cite{rava}. 

Equation\ (\ref{eq1}) represents the probability that an element fails under the 
individual load $\sigma$. Now, consider the case in which an element drawn from Eq.\ 
(\ref{eq1}) supports a load $\sigma_1$, but breaks under a new load $\sigma_2$. The 
probability that this happens is given by 
 \begin{equation}
 p(\sigma_1,\sigma_2)=\frac{P(\sigma_2)-P(\sigma_1)}{1-P(\sigma_1)},
 \label{eq2}
 \end{equation}
which is equal to
 \begin{equation}
 p(\sigma_1,\sigma_2)=1-e^{-(\sigma_2^{\rho}-\sigma_1^{\rho})}.
  \label{eq3}
 \end{equation}

So, the probability $q(\sigma_1,\sigma_2)$ that an element that
has survived to the load $\sigma_1$ also survives to the load
$\sigma_2$ will be given by
 \begin{equation}
 q(\sigma_1,\sigma_2)=1-p(\sigma_1,\sigma_2)=e^{-(\sigma_2^{\rho}-\sigma_1^{\rho})}.
 \label{eq4}
 \end{equation}

All these probabilities depend on the state of stress of each element, which is a complex 
function of both the control parameter and the stress redistributions due to fiber 
failures. To mimic the quasistatic increase in load on the system as is applied in MC 
simulations, we impose the condition that under an external force $F$, the next breaking 
event consists of one single failure. Now, we will explore how the driving process 
controls the approach to global failure in the global load sharing case and in a 
particular version of the local load sharing schemes. 

\subsection{Global Load Sharing case} \label{ELScase} 

Let suppose that after the latest avalanche, there are $N_k$ surviving elements each one 
bearing a load $\sigma_k$. The assumption of quasistatic increase in
load on the system implies that the load $\sigma_l$ needed to provoke
the failure of only one element is \cite{mgpprl00},
 \begin{equation}
 \sigma_l=\left[\sigma_k^{\rho}-\ln\left(1-\frac{1}{N_k}\right)\right]^{\frac{1}{\rho}},
 \label{eq6}
 \end{equation}
where in Eq.\ (\ref{eq6}) $N_k=N_0$ and $\sigma_k=0$ at the initial state. Elevating the 
external force up to the $N_k\cdot\sigma_l$ level a first element breaks. As we are 
dealing with a global load sharing set, the choice of the broken element is irrelevant 
because all of them are equivalent. Once the first element fails, the redistribution of 
its stress takes place which may induce secondary failures and so on, until the end of 
the avalanche. 

Now, the number of surviving elements in the set after an avalanche
has come to an end can be calculated by the recursive relations \cite{mgpprl00},
 \begin{equation}
 N_{j+1}=N_j\cdot q(\sigma_{j-1},\sigma_{j}),
 \label{eq8}
 \end{equation}
 \begin{equation}
 N_j\cdot\sigma_j=N_{j-1}\cdot\sigma_{j-1}
 \label{eq9}
 \end{equation}
\begin{equation}
 N_j=N_{j+1}
 \label{eq10}
 \end{equation}

Equation\ (\ref{eq8}) relates the number of surviving elements between
two successive avalanche steps, whereas Eq.\ (\ref{eq9}) allows the
computation of the new load acting on the $N_j$ surviving
elements. This is possible because we are under a global load sharing
scheme and thus all surviving elements support the same load. Finally,
Eq.\ (\ref{eq10}) determines the end of the ongoing avalanche.

The dynamics of the system is completely determined by Eq.\ (\ref{eq6}), (\ref{eq8}), 
(\ref{eq9}). In this way, the size of an avalanche is given by the number of elements 
that break between two successive steps of external loading. The total stress accumulated 
in the system can be calculated multiplying the number of intact fibers before an 
avalanche starts by the load given by Eq.\ (\ref{eq6}). The critical load, defined as the 
load needed to provoke the total collapse of the system, is equal to the load on the 
intact fibers just before the final catastrophic avalanche. Note that in this 
probabilistic approach, in contrast to Monte Carlo simulations, we need to store only the 
information concerning the loads of the intact elements, that is, the threshold dynamics 
is omitted with the subsequent advantages of saving computer resources and the 
possibility of exploring systems of larger size. 

For Eq.\ (\ref{eq10}) to be satisfied, we can proceed in two different ways in order to 
determine when an avalanche ends, to which we will refer as the continuous and the 
discrete cases. For the continuous case, the number $N_{j+1}$ of surviving elements is 
considered a real number. This means that condition\ (\ref{eq10}) is never fulfilled 
before the final avalanche. So, condition\ (\ref{eq10}) is replaced in numerical 
simulations by a factor $\nu\ll1$ that determines the end of an avalanche, i.e., if 
$N_{j}-N_{j+1}\leq\nu$ the avalanche stops; otherwise it continues. In the discrete case, 
$N_{j+1}$ is considered to be an integer number, so that after each iteration of Eq.\ 
(\ref{eq8}), $N_{j+1}$ has to be rounded up. This is done comparing the remainder of 
$N_{j+1}$, $\lambda$, with a random number $\alpha$ uniformly distributed in the interval 
$[0,1[$. Thus, if $\alpha\geq\lambda$, $N_{j+1}$ is equal to its whole part, and if not, 
$N_{j+1}$ is equal to its whole part plus one . Next, we check whether condition\ 
(\ref{eq10}) is satisfied for the rounded value of $N_{j+1}$ or if a new iteration of 
Eq.\ (\ref{eq8}) has to be performed.  The continuous approach has 
the great advantage that the fluctuations are ruled out, whereas for the discrete case 
the results are similar to those obtained by Monte Carlo simulations where it is 
necessary to average over many realizations in order to get accurate mean values. This is 
because in the global fiber bundle model the central limit theorem applies \cite{s89}. 

Now, we proceed to explore the behavior of the system near the critical point. In 
particular, we are interested in inspecting the evolution of some quantities as the 
critical point is reached. In order to avoid unnecessary fluctuations around the critical 
point, the results shown below have been obtained for the continuous case of the 
probabilistic approach ($\rho=2$). It turns out \cite{mgpprl00} that the avalanche size near to the 
critical point diverges with an exponent $\gamma=\frac{1}{2}$ as 
$s\sim(\sigma_c-\sigma)^{-\gamma}$. A similar behavior, through a mapping of a fuse 
network model to the global fiber bundle model used here, has been recently reported 
\cite{zrsv97}. This mapping between fiber bundle models and fuse networks with strong 
disorder was first noted \cite{hhphyleta94} a few years ago; nevertheless, it has also 
been pointed out that 3D fuse networks apparently do not follow the FBM picture 
\cite{miko}. Note, additionally, that for the probabilistic continuous version, this 
relation is obtained for a single realization avoiding, in this way, the large number of 
iterations performed in MC simulations. In Fig.\ \ref{fig1} we have
depicted the derivative of the number of broken 
fibers, ($dN/d\sigma$), as a function of the distance to the critical point 
$\sigma_c-\sigma$ in a log-log plot. This rate $dN/d\sigma$ diverges as 
$(\sigma_c-\sigma)^{-\gamma}$ also with $\gamma=\frac{1}{2}$, thus qualifying a critical 
mean field behavior as was already shown \cite{asl97} by means of analytical analysis of 
fiber bundle models. In Ref.\ \cite{kzh99}, a similar scaling behavior is addressed for 
the derivative of the strain carried by the fibers with respect to the driving force. 

Another way to shed light on the critical behavior of this type of system is to define a 
branching ratio $\zeta$ for each avalanche. This magnitude represents the probability to 
trigger future breaking events given an initial individual failure \cite{jensen,ccp99}, 
and is related to the number of broken fibers by 
 \begin{equation}
 \zeta=\frac{<z>-1}{<z>}.
 \label{eq11}
 \end{equation}
The above relation can be obtained by thinking of the evolution of fracture as a kind of 
branching process \cite{h63}. In this process, each node gives rise to a number $n$ of 
new branches in the next time step. The average number $<n>$ of new branches is called 
the branching ratio. Let us denote by $n_t$ the number of branches at a given step $t$ of 
the branching process, and by $t_{max}$ the total number of time steps before it dies. 
Then, 
 \[
\zeta=<n>=\frac{\sum\limits_{t=0}^{t_{max}-1}n_{t+1}}{\sum\limits_{t=0}^{t_{max}}n_t},
 \]
and
 \[
\zeta=1-\frac{n_0}{\sum\limits_{t=0}^{t_{max}}n_t}. 
 \]

As $n_0=1$, $\zeta=1-\frac{1}{n_{tot}}$ where $n_{tot}$ is the total number of nodes 
developed in the branching process. For a fracture process, $n_{tot}$ is equal to the 
average number of failure events. So, Eq.\ (\ref{eq11}) defines the branching ratio. We 
represent by $<z>$ the average number of elements that fail in one avalanche, which is a 
function of the control parameter $\sigma$ and coincides with $s$. This analogy between 
fracture and branching processes has been previously used to study the criticality in the 
process of fragmentation of Hg drops \cite{smla96}. The branching ratio will then act as 
the order parameter. It takes the value 1 when the system is critical thereby 
representing a measure of the distance of the system from the critical state 
\cite{ccp99}. We would like to remark here two characteristics of the branching ratio 
defined as above. First, it coincides with its general definition for a branching 
process. Here, the values of the branching ratio are always less than or equal to one 
because we are analyzing an irreversible breaking process that cannot continue forever, 
so that a value of the branching ratio greater than one would imply a physically 
unreachable situation. Secondly, we have introduced a definition in terms of the {\em 
avalanche size} and not in terms of the {\em cluster size}. The avalanche size in fiber 
bundle models is a measure of {\em causally} connected broken sites while the cluster 
size is a measure of {\em spatially} connected broken sites. It is clear that in the 
global fiber bundle model, the spatial correlations are ruled out and then the 
distribution of avalanche sizes does not coincide with the distribution of cluster sizes. 

We have numerically computed $\zeta$ for the continuous version of the probabilistic 
model. The results obtained for a system of $N_0=50000$ elements and $\rho=2$ have been 
plotted in Fig.\ \ref{fig2}. It can be seen in this figure that the branching ratio 
approaches the unity as the critical load is reached. In the figure, the values of 
$\zeta$ are collected for all the avalanches except for that which provokes the 
catastrophic event leading to the collapse of the system. So, this dependency of $\zeta$ 
with $\sigma$ means that very close to the critical point, the initial failure provokes 
the breaking of another fiber which in turn induces tertiary ruptures and so on. This is 
the result that should be expected from the divergence of the avalanche size at the 
critical point. Figure\ \ref{fig3} shows the behavior of the order
parameter near the critical point. As can be seen, near the critical point, the relation $1-\zeta \sim 
(\sigma_c-\sigma)^{\beta}$, where $\beta=\frac{1}{2}$ applies. Note the similarity of the 
behavior with those obtained for the magnetization in known magnetic systems with 
second-order phase transitions. It is of additional interest to note that the branching ratio also captures the feature that the precursory 
activity is significant only for strong disorder (for the Weibull distribution, small 
values of $\rho$); whereas for small disorder the system behaves more similar to the 
breakdown of homogeneous materials. On the other hand, the branching ratio does not 
depend on the size of the system for large systems, in contrast to previous results in 
other fracturing systems \cite{ccp99}. In Fig.\ \ref{fig4} we illustrate this behavior by 
plotting the value of the branching ratio for the last non-catastrophic avalanche, 
$\zeta^{*}$, as a function of the system size. In all cases, the numerical simulations 
were performed for $\rho=2$. It is clear from the figure that
$\zeta^{*}$ reaches the unity as $N_0$ goes to infinity.  

\subsection{Local Load Sharing Case}
\label{LLScase}

It is well-known that the local load sharing FBM is much more complicated than the global 
load sharing case. The complexity of the fracture problem increases because the load 
borne by failed elements is transferred to nearest neighbors and then there appear 
regions of stress concentration throughout the system. The distribution of load is now 
not homogeneous, and we have to carry on several lists to record the individual load of 
the fibers in the system. However, with the probabilistic approach, we are able to study 
a few things about the behavior of the system, which are sufficient to remark the great 
differences between the long range and the short range interactions schemes. One 
important difference between the two schemes is that contrary to the global load sharing 
case, in the local scheme there is no significant precursory activity signaling the 
approach to the final collapse and the system undergoes an abrupt catastrophic avalanche. 
For maximum simplicity, we consider next a one-dimensional periodic local load system 
where the stress transfer is done by adding the load of the failed element always to the 
element on its right. Despite of its simplicity, it has been shown \cite{gip93} that the 
general properties of this particular model are identical to those of more complex local 
schemes. 

The probability that the system fails in only one avalanche is given
by:
 \begin{equation}
 p(\sigma_1,\sigma_2)\cdot p(\sigma_1,\sigma_3)\cdot p(\sigma_1,\sigma_4)\ldots
 p(\sigma_1,\sigma_{N_0}),
 \label{eq12}
 \end{equation}
where $\sigma_i=i\sigma_1$ and $\sigma_1$ is 
 \[ 
\sigma_1=\left(0-\ln\left(1-\frac{1}{N_0}\right)\right)^{\frac{1}{\rho}}.
 \]

That is, the probability of having a one-step failure is given by the probability that an 
element fails under a load $\sigma_2$ having survived to the load $\sigma_1$, multiplied 
by the probability that the next element in the lattice also fails under the new load 
$\sigma_3$ having survived to the load $\sigma_1$ and so on. 

On the other hand, Eq.\ (\ref{eq4}) takes the form
 \begin{equation}
 q(i,j)=e^{-\sigma_1^{\rho}(j^{\rho}-i^{\rho})},
 \label{eq13}
 \end{equation}
which gives the probability that an element that has survived to the load 
$i\cdot\sigma_1$ also survives to a new load $j\cdot\sigma_1$, being $\sigma_1$ the load 
that produces the first breaking event in the system at the initial state. The 
probability that the first avalanche consists of, for example in one failure, is given by 
$q(1,2)$. The first avalanche will have size two with probability $p(1,2)\cdot q(1,3)$. 
In general, the probability that the first avalanche does not provoke the rupture of the 
whole system, i.e., be finite, is given by 
 \begin{equation}
 P_1=\sum_{l=1}^{\infty}\Pi_l\cdot q(1,l+1),
 \label{eq14}
 \end{equation}
where
 \[
 \Pi_l=p(1,2)\cdot p(1,3)\ldots p(1,l)=\prod_{i=2}^{l}p(1,i),\qquad
 l<N_0,
 \]
with $\Pi_1=1$. Thus, the average size of the first avalanche,
derived from Eq.\ (\ref{eq14}) is
 \begin{equation}
 S_1=\sum_{l=1}^{\infty}l\cdot\Pi_l\cdot q(1,l+1).
 \label{eq15}
 \end{equation}
Numerical simulations of the continuous case, the discrete approximation (both defined as 
above for the global load sharing scheme) and Monte Carlo method confirm the validity of 
Eq.\ (\ref{eq15}). For example, for $\rho=2$ and $N_0=1000$ we obtain $S_1=1.0030$, 
$S_1=1.0022$ and $S_1=1.0039$ respectively. This result is valid only for the first 
avalanche in the system. 

We have also verified by means of Monte Carlo simulations that for this local transfer 
rule, there is no power law in the distribution of avalanche sizes. Besides, no scaling 
relations appear and the system is more sensitive to parameters such as the degree of 
disorder ($\rho$) and the size of the system $N_0$. Catastrophic avalanches arise very 
often in the first stages of the rupture process. 

Finally, we illustrate how one can describe the precursory activity when it is limited to 
a few steps before the global rupture. Let us solve the particular situation in which the 
first avalanche is finite and the second provokes the final breakdown of the system. The 
load needed to break the first element at the initial state is
 \[ 
\sigma_1=\left(-\ln\left(1-\frac{1}{N_0}\right)\right)^{\frac{1}{\rho}}\simeq\left(\frac{1}{N_0}\right)^{\frac{1}{\rho}}, 
\qquad N_0\gg 1.
 \]
 So $\sigma_1^{\rho}\simeq\frac{1}{N_0}$, and 
 \begin{equation}
 q(1,2)=e^{-(2^{\rho}-1)\cdot\frac{1}{N_0}}.
 \label{eq16}
 \end{equation}
If the value of $q(1,2)$ in Eq.\ (\ref{eq16}) is bigger than $0.5$, then the fracture 
will likely consist of one single event of rupture and the avalanche will not progress 
beyond this first failure. As shown above, for $\rho=2$ a system size of $N_0=1000$ 
elements is enough to produce first avalanches of size $1$ on average. Next, the driving 
force is increased and the new load acting on the surviving $N_0-1$ fibers is \[ 
\sigma_2=\left(\sigma_1^{\rho}-\ln\left(1-\frac{1}{N_0-1}\right)\right)^{\frac{1}{\rho}}\simeq\left(\frac{2}{N_0}\right)^{\frac{1}{\rho}}, 
\qquad N_0-1\gg 1 \] that is, $\sigma_2^{\rho}=\frac{2}{N_0}$. In the derivation of 
$\sigma_2$ it has been assumed that the second avalanche starts with high probability 
in a fiber different of that which is located just in the crack tip of the first 
avalanche. This approximation is well-justified by numerical simulations of this local 
load model for which it is known that the first breaking events are randomly distributed 
in the sample, leading to the appearance of several small cracks which much later 
coalesce and grow provoking the final breakdown. Now, there are $(N_0-2)$ elements 
bearing a load $\sigma_2$ and one element supporting a load $\sigma_2+\sigma_1$. The 
system will experience on average the final breakdown in this second breaking event if 
the probability 
 \begin{equation}
 q'(1,2)=e^{-(2^{\rho}-1)\cdot\frac{2}{N_0}}
 \label{eq17}
 \end{equation}
is lower than $0.5$. Nevertheless, $q(1,2)$ and $q'(1,2)$ are not
independent since the relation
 \begin{equation}
 q'(1,2)=\left(q(1,2)\right)^{2}
 \label{eq18}
 \end{equation}
has to be verified. This dependence, together with the
restrictions $q(1,2)>0.5$ and $q'(1,2)<0.5$ leads to
$0.5<q(1,2)<0.7$ and $0.25<q'(1,2)<0.5$ as acceptable values for
$q(1,2)$ and $q'(1,2)$ in order to have a two-step system
collapse. This approximation seems reasonable taking into account
the way in which numerical simulations of the continuous and the
discrete versions of the probabilistic approach proceed. So, the
condition for the first avalanche to be finite and that the second
breaks the system is reduced to
 \begin{equation}
 N_0\simeq2^{\rho+1}-2
 \label{eq19}
 \end{equation}
for $q(1,2)=0.6$, a value in the middle of its range. This relationship between $N_0$ and 
$\rho$ is confirmed in numerical simulations. The situation considered above corresponds 
to a very brittle rupture since the ratio of the size of the typical local damage that 
induces the system failure to the system size is very close to one $(N_0-1)/N_0$. 
Expression\ (\ref{eq19}), although very simple, gives a rough estimate of the qualitative 
relation between the size of the system and the amount of disorder when the degree of 
brittleness of the fracture is high. The situation analyzed becomes important and more 
accurate as we go to larger system sizes because for the local load transfer rule and a 
larger number $N$ of fibers, the probability to find a weak region somewhere in the 
system is high and due to the local redistribution rule the crack will propagate fast 
enough to break the whole system in a few steps (strictly speaking, in the thermodynamic 
limit the critical load is zero). 

\section{Discussion and Conclusions}
\label{discon}

We have proposed a probabilistic approach to fiber bundle models of fracture. The cases 
of long range interactions among the fibers of the bundle and a local load sharing 
transfer scheme (short range interactions) were considered in order to investigate the 
behavior of the system near the global breakdown of the material. The results obtained 
for the local scheme indicate that the system undergoes a kind of first-order phase 
transition in agreement with previous reports \cite{zrsv97,wl99}. The system fails with 
no significant precursors announcing the incipient rupture, and no scaling relations can 
be found even for large system sizes and strong disorder. We illustrated this behavior by 
considering the situation of a two-step global failure. In this particular case, several 
quantities have a discrete jump at the critical point, like the branching ratio $\zeta$ 
which goes from zero to $\frac{N_0-2}{N_0-1}$, i.e., from zero to one in the limit of 
$N_0\rightarrow\infty$. 

The scenario is quite different for the global load sharing FMB. In this case, the type 
of phase transition is not very clear. There are basically two possible ways in which 
such a transition may occur. In a first-order phase transition, we find discontinuous 
behavior in various quantities as we pass through the critical point. Contrary to this 
case, in a second-order (or continuous) phase transition, the fluctuations are correlated 
over all length scales, therefore, not only does the correlation length diverges 
continuously as the critical point is approached, but also other quantities show scaling 
\cite{c96}. A particular case is the first-order phase transition close to a 
spinodal-like instability, for which scaling relations can be obtained despite that some 
macroscopic quantities, like the elastic modulus, have a discrete, finite jump at the 
critical point \cite{m94}. The question then is whether the global load sharing FBM is a 
case of a second-order transition or a first-order spinodal transition. 

The results obtained with the probabilistic approach seem not to support the claim of 
Ref.\cite{zrsv97} that the failure of a fiber bundle model under a global load sharing 
transfer scheme behaves as a first-order phase transition close to a spinodal point. 
There, by simulating models of electric breakdown and fracture, the authors presented 
ample numerical and theoretical evidence of several scaling relations and of the discrete 
jump of the macroscopic properties. We have obtained the same scaling relation for the 
rate of fiber failures as the critical point is reached, as well as for the avalanche 
sizes, which also diverge at the transition. Note that the scaling exponent derived from numerical simulations of the 
continuous version of the probabilistic approach fits very well the mean field result 
$\gamma=\frac{1}{2}$. Besides, the fraction of unbroken fibers just before the global 
rupture has a discontinuity, which is not size dependent for big systems as can be 
observed in Fig.\ \ref{fig5}. It has also been shown by analytic means
that the distribution of avalanche sizes in the case of global load sharing transfer rule follows 
a universal power law with an exponent $-5/2$ \cite{hh92,khh97,s289}. However, we should 
notice that in driven disordered systems, the concepts related to spinodal nucleation are 
not sufficiently well established. In particular, it seems that the finite size of the 
system makes it possible that the last catastrophic avalanche provokes the elastic 
modulus to drop to zero with a finite, and very likely, N-dependent jump. Thus, from our 
point of view, that is not enough to set the conclusion that fracture can be described as 
a first-order phase transition.

Our alternative point of view is to consider the {\sl branching ratio} defined previously 
as a measure of the distance of the system from the critical point. According to the 
results obtained, the branching ratio goes {\sl continuously} from zero to one. The branching ratio measures the probability that a fiber that has failed 
triggers none, one or more breaking events. Therefore, a branching ratio equal to one at 
the critical point implies that, on average, one failure will induce another failure so 
that the system reaches a state where any perturbation propagates across the entire 
system, which is, in essence, a critical phase. This invokes a continuous phase 
transition as claimed in other analysis of fracture models \cite{sa98}. Note, 
additionally, that what changes discontinuously at $\sigma_c$ is the rate of change of 
$\zeta$ rather than $\zeta$ itself. 

In summary, we have introduced a probabilistic approach to fiber bundle models which 
allows to avoid the fluctuations near the critical point. The different rupture behaviors 
of the system for the global load sharing scheme and the local one were shown. For the 
short-range interactions case, the rupture is no-doubt of the first-order phase 
transition type. For the global transfer rule, several scaling relations were obtained 
with mean field critical exponents. The branching ratio was defined as an appropriate 
order parameter. According to the results obtained, the branching ratio goes continuously 
from zero to one. This suggests that fracture in heterogeneous systems with long range 
interactions can be described as a phase transition of the second-order type, at least 
within the FBM picture. 

\section{ACKNOWLEDGMENTS}

Y.M thanks A. Vespignani and H. J. Jensen for very useful and stimulating discussions. Y. 
M also thanks The Abdus Salam International Centre for Theoretical Physics for 
hospitality and financial support. This work was supported in part by the Spanish DGICYT 
under Project PB98-1594.

\newpage

\begin{figure}
 \caption{Rate of fiber failure as a function of the distance to the
critical point for the continuous version of the probabilistic global load sharing model 
for a system of $N_0=50000$ elements and $\rho=2$. A line with the mean-field value 
$\gamma=\frac{1}{2}$ of the exponent is plotted for reference.} 
 \label{fig1}
\end{figure}

\begin{figure}
\caption{Evolution of the branching ratio as the critical point is approached in the 
continuous probabilistic method ($N_0=50000$). Note that at the critical point the 
branching ratio reaches the unity.} 
 \label{fig2}
\end{figure}

\begin{figure}
\caption{Behavior of the branching ratio near the critical point. The
straight line satisfies the scaling relation
$(1-\zeta)\sim(\sigma_c-\sigma)^{\beta}$, with $\beta=0.5$. This plot
has been obtained for a uniform distribution of fiber's strengths.}
\label{fig3}
\end{figure}

\begin{figure}
\caption{Branching ratio for different system sizes ($\rho=2$) for the global load sharing 
case. As the size of the system is increased the value of the branching ratio for the 
last non-catastrophic avalanche approaches unity.} 
 \label{fig4}
\end{figure}

\begin{figure}
\caption{Fraction of unbroken fibers just before the final breakdown in the probabilistic 
model of the global load sharing case as a function of the system size ($\rho=2$). The 
horizontal line shows the exact value in the thermodynamic limit.} 
 \label{fig5}
\end{figure}

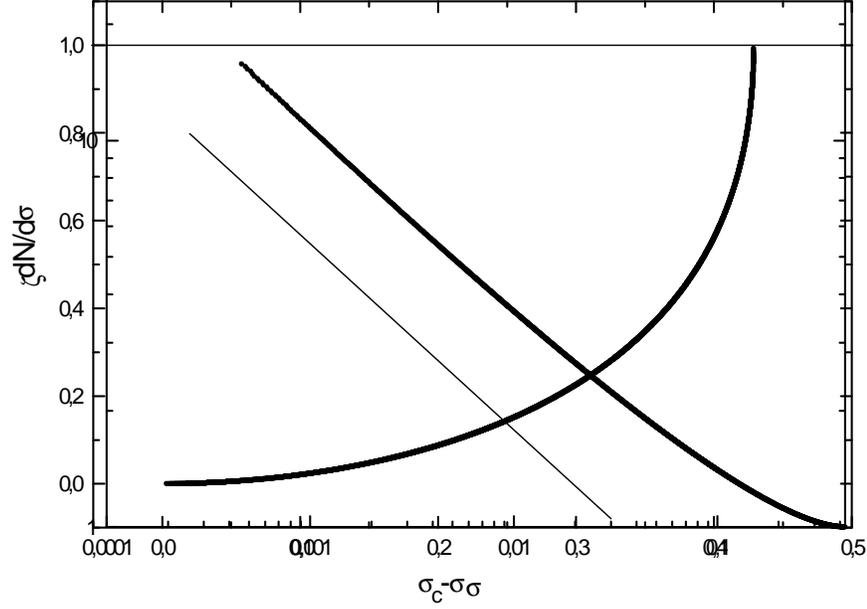
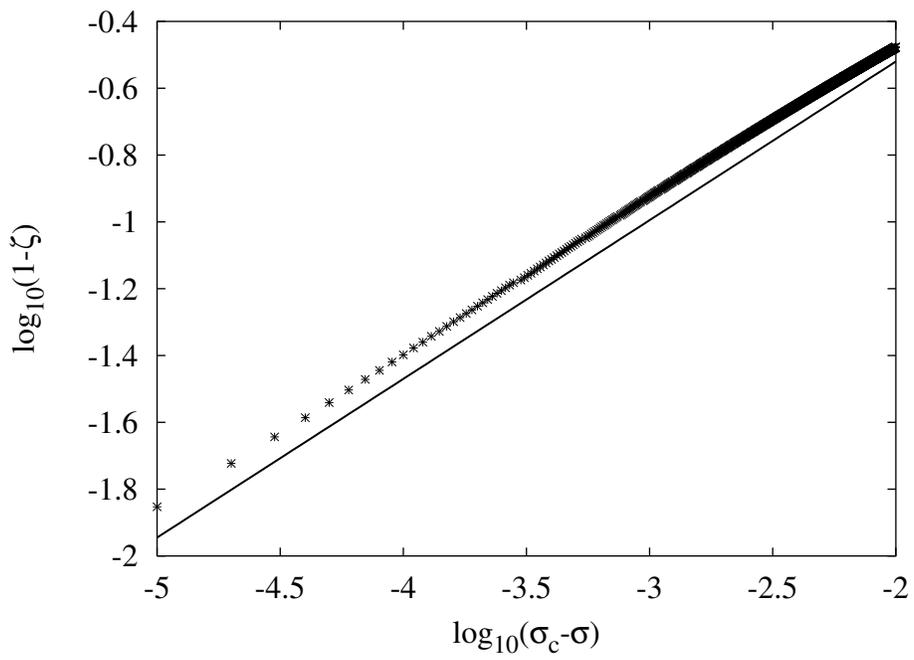
\end{document}